\documentclass[aps,twocolumn,showpacs,amsmath,amssymb]{revtex4}

\usepackage[english]{babel}
\usepackage{graphicx}

\begin{document}

\newcommand{\unit}[1]{\:\mathrm{#1}}            
\newcommand{\To}{\mathrm{T_0}}
\newcommand{\Tp}{\mathrm{T_+}}
\newcommand{\Tm}{\mathrm{T_-}}
\newcommand{\EST}{E_{\mathrm{ST}}}
\newcommand{\Rp}{\mathrm{R_{+}}}
\newcommand{\Rm}{\mathrm{R_{-}}}
\newcommand{\Rpp}{\mathrm{R_{++}}}
\newcommand{\Rmm}{\mathrm{R_{--}}}

\bibliographystyle{unsrt}
\title{Decoherence-avoiding spin qubits in optically active quantum dot molecules}
\author{K. M. Weiss$^{\dagger}$}
\author{J. M. Elzerman$^{\dagger}$}
\author{Y. L. Delley}
\author{J. Miguel-Sanchez}
\author{A. Imamo\u{g}lu}
\affiliation{Institute of Quantum Electronics, ETH Zurich, CH-8093
Zurich, Switzerland.\\
$^\dagger$These authors contributed equally to this work.}

\begin{abstract} 
In semiconductors, the $T_2^*$ coherence time of a single confined spin is limited either by the fluctuating magnetic environment (via the hyperfine interaction), or by charge fluctuations (via the spin-orbit interaction). We demonstrate that both limitations can be overcome simultaneously by using two exchange-coupled electron spins that realize a single decoherence-avoiding qubit. Using coherent population trapping, we generate a coherent superposition of the singlet and triplet states of an optically active quantum-dot molecule, and show that the corresponding $T_2^*$ may exceed 200 nanoseconds.
\end{abstract}


\maketitle
A single spin-$1/2$ particle such as an electron represents the prototypical two-level quantum system. However, its simplicity leaves no room for designing the energy levels to be robust against various sources of decoherence. This means single-spin coherence~\cite{Hanson:2007,Mikkelsen:2007,Press:2008,Xu:2008,Brunner:2009,DeGreve:2011} can only be improved by dynamically decoupling the spin from its environment using echo techniques~\cite{DeGreve:2011,Petta:2005,Bluhm:2010,Press:2010}, or by reducing environmental fluctuations~\cite{Latta:2009,Xu:2009,BluhmPRL:2010}. To bypass such elaborate procedures, we have to go beyond single-spin states and build robustness against decoherence directly into the energy spectrum~\cite{Lidar:1998}. In particular, coupling two electron~\cite{Kim:2010,Elzerman:2011} or hole~\cite{Greilich:2011} spins via a strong exchange interaction rewards us with a tunable energy spectrum that exhibits entangled spin singlet and triplet states~\cite{Stinaff:2006,Tureci:2007}. The coupled system features a ``sweet spot'' in the bias parameters, where the qubit subspace spanned by the singlet state (S) and the triplet state with spin z-projection $\mathrm{m_s}=0$ ($\To$) is to first order insensitive to magnetic as well as electric field fluctuations, similar to what has been achieved in superconducting quantum circuits~\cite{Vion:2002,Koch:2007}. 

\begin{figure}[t]
\includegraphics[width=75mm]{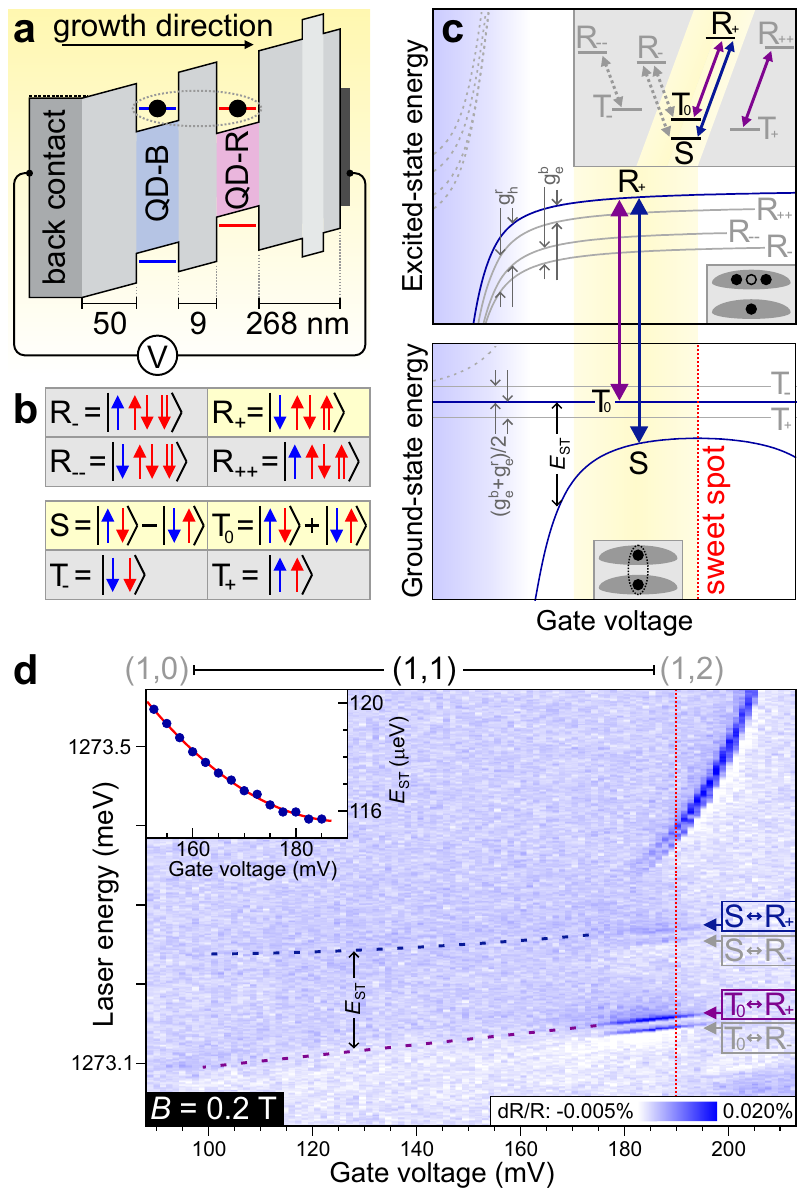}
\caption{(color online). (a) Schematic energy diagram of the device, containing two layers of self-assembled InGaAs QDs, separated by a $9\unit{nm}$ GaAs tunnel barrier and embedded in a GaAs Schottky diode. (b) Ground and lowest-lying optically excited states in the (1,1) regime. Blue arrows indicate electron spins in the bottom QD, red single (double) arrows indicate electron (hole) spins in the top QD. (c) Schematic energy diagram of the ground and optically excited states versus $V$. A magnetic field in Faraday geometry induces Zeeman splittings proportional to the g-factors shown, with $g^r_e$ ($g^b_e$) denoting the electronic g-factor in the red (blue) QD, and $g^r_h$ the hole g-factor in the red QD. The dashed red line indicates the sweet spot, where $d\EST/dV = 0$. Inset: Circularly polarized dipole-allowed optical transitions between the states shown in (b). (d) Differential reflection (dR) measurement of the trion transitions in the red QD of CQD1 versus $V$ at $B=0.2~\unit{T}$, measured around saturation power (laser Rabi frequency $\Omega=0.8~\unit{\mu eV}$) in the presence of a weak non-resonant ($850~\unit{nm}$) laser. Blue (purple) dashed lines indicate the $\mathrm{S}-\Rp$ ($\To-\Rp$) transition energies, extracted from two-laser repump measurements~\cite{supplemental}. The unmarked diagonal feature in the top right-hand corner is due to indirect transitions involving the (1,2) charging ground state. Inset: $\EST$ versus $V$, including a parabolic fit (red line).
}
\label{Scheme}
\end{figure}

Although ground breaking experiments based on S-$\To$ states have been carried out using electrically-defined coupled quantum dots (QDs), these were operated far from the sweet spot and with a minimal exchange coupling much smaller than the Overhauser field gradient~\cite{Hanson:2007,Petta:2005,Bluhm:2010,BluhmPRL:2010}. Therefore the S-$\To$ qubit was always exposed to either magnetic or electric field fluctuations. In optically active QD molecules, on the other hand, experiments have been performed in the large-exchange regime, but still away from the
sweet spot~\cite{Kim:2010,Greilich:2011}. Here, we demonstrate for the first time that operation at the sweet spot is indeed a promising strategy, prolonging the $T_2^*$ coherence time by two orders of magnitude. In fact this system could be considered as a solid-state analog of atomic clock states~\cite{Vanier:2005}: the possibility of optical Raman coupling  between the two clock states (S and $\To$) allows for manipulation of the qubit at the sweet spot where all unwanted low-frequency couplings vanish, ensuring full protection from noise. 

Our experiments utilize a pair of tunnel-coupled self-assembled InGaAs QDs~\cite{Stinaff:2006,Krenner:2005}. By adjusting the growth parameters (Fig.~\ref{Scheme}a) we ensure that both QDs are charged with a single electron for a wide range of the applied gate voltage $V$. In this so-called (1,1) regime~\cite{Kim:2010,Elzerman:2011,Greilich:2011} the S and $\To$ ground states (Fig.~\ref{Scheme}b) are split by a voltage-dependent exchange interaction $\EST$ (see the lower panel in Fig.~\ref{Scheme}c). For a particular gate voltage $V_0$ (the sweet spot), $d\EST/dV = 0$ so that $\EST$ is to first order insensitive to electric-field fluctuations. In addition, the large value of $\EST$ suppresses mixing between the S and $\To$ states arising from the Overhauser field gradient. Finally, hyperfine mixing between the three triplets (T) is suppressed by applying an external magnetic field ($B$) along the growth direction $z$, which splits off $\mathrm{T_{\pm}}$ (with spin z-projection $\mathrm{m_s}=\pm 1$), while leaving both S and $\To$ unaffected. Under these conditions the two-level system of S and $\To$ is therefore extremely robust against both charge and nuclear-spin fluctuations and forms a decoherence-free subspace~\cite{Lidar:1998}. To demonstrate this experimentally, we focus on the lambda system formed by S and $\To$ plus the shared optically excited state $\mathrm{R_+}$ that contains a trion in the red QD (see the upper panel in Fig.~\ref{Scheme}c). 

\begin{figure}[t]
\includegraphics[width=68mm]{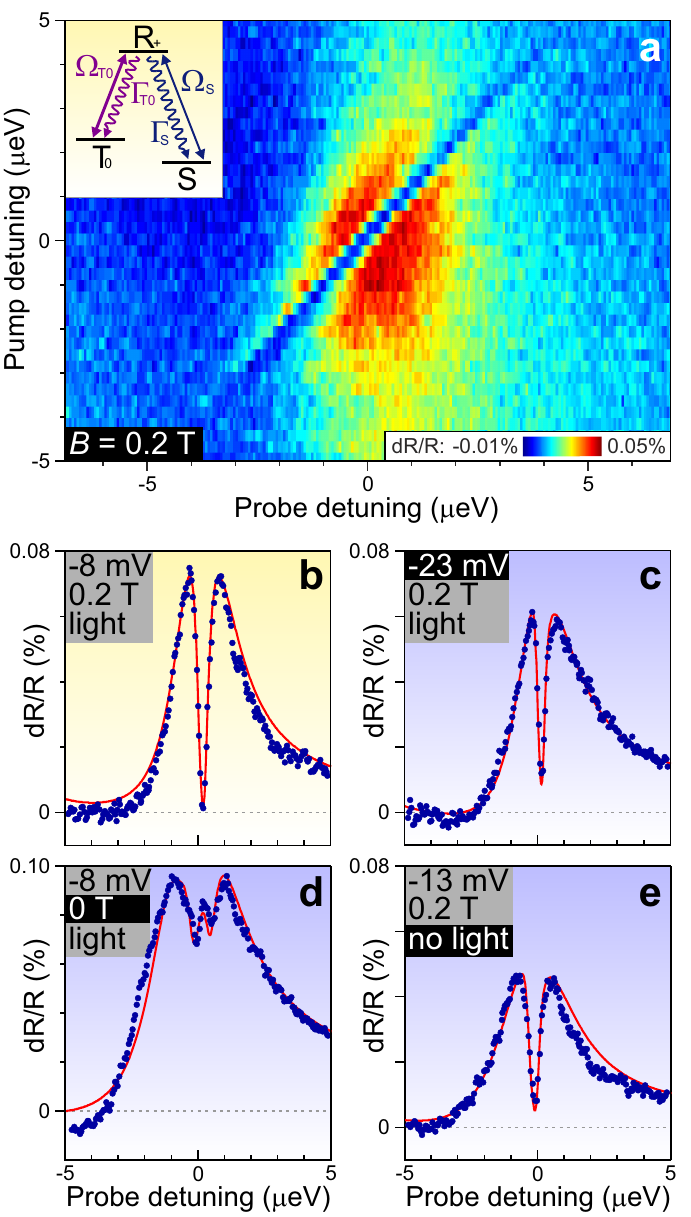}
\caption{(color online). Coherent population trapping with a ST qubit (CQD1). (a) CPT measured in dR versus pump and probe laser detuning  at $\Delta V=-13~\unit{mV}$ and $B=0.2~\unit{T}$. The probe laser ($\Omega_{\mathrm{S}}=0.34~\unit{\mu eV}$; incident on the $\mathrm{S}-\Rp$ transition) and the pump laser ($\Omega_{\To}=0.83~\unit{\mu eV}$; incident on the $\To-\Rp$ transition) have orthogonal linear polarization. Inset: Schematic diagram of the right-hand circularly polarized lambda scheme. (b) dR (blue dots) versus probe detuning at $B=0.2~\unit{T}$ and $\Omega_{\To}=0.77~\unit{\mu eV}$, in the presence of a weak non-resonant ($850~\unit{nm}$) laser that reduces the charge fluctuations. The red line is a numerical fit to an eight-level model as described in the supplemental material~$(27)$. (c) Same as in (b), but with $\Omega_{\mathrm{T_0}}=0.58~\unit{\mu eV}$ and $\Delta V=-23~\unit{mV}$. (d) Same as in (b), but at $B=0~\unit{T}$. (e) Same as in (b), but at $\Delta V=-13~\unit{mV}$ and without the non-resonant laser.
}
\label{CPT}
\end{figure}

We employ single-laser differential reflection (dR) measurements~\cite{Alen:2003,supplemental} to  map out the optical transitions of the red QD versus $V$. In the (1,1) regime we observe very efficient spin pumping into the S (T) state while probing the T (S) transitions, as evidenced by a vanishing dR contrast (Fig.~\ref{Scheme}d). All transitions driven by a single laser field are only visible in a narrow gate-voltage range at the edge of the (1,1) regime, where spin-flip tunneling processes to and from the back contact lead to spin relaxation between the ground states~\cite{Elzerman:2011}. By having a resonant laser present on both the $\mathrm{S-R_{+}}$ and the $\mathrm{T_0-R_{+}}$ transition simultaneously, the spin pumping is lifted and we can determine the voltage dependence of $\EST$ (inset to Fig.~\ref{Scheme}d). From this we find the sweet spot to be at $V_0=190~\unit{mV}$, just outside the (1,1) regime for this particular QD molecule (which we call CQD1).

In order to measure the coherence properties of the two-level system formed by S and $\To$, we rely on the quantum optical technique of coherent population trapping (CPT)~\cite{Fleischhauer:2005,Boller:1991}. A weak probe laser is
tuned across the $\mathrm{S}-\Rp$ transition while a non-perturbative coupling laser is incident on the $\To-\Rp$ transition (see the inset to Fig.~\ref{CPT}a). At the two-photon resonance the QD molecule is prepared in an optically dark state consisting of an antisymmetric superposition of S and $\To$. Here destructive interference between the two optical transition paths leads to a vanishing photon scattering amplitude, and thus a dip (or \emph {dark-resonance}) in the dR spectrum (Fig.~\ref{CPT}a). Since this transparency results from the formation of a coherent superposition of S and $\To$, decoherence processes with both slow and fast decorrelation times lead to a suppression of the CPT dip.

We observe that the $\mathrm{S}-\To$ coherence  is highly sensitive to the external magnetic field and the applied gate voltage. At $B=0.2~\unit{T}$ and $\Delta V = V-V_0=-8~\unit{mV}$, the CPT dip goes completely to zero for a pump laser Rabi frequency of $\Omega_{T0}=0.79~\unit{\mu eV}$ (Fig.~\ref{CPT}b). Tuning $V$ away from the sweet spot to $\Delta V=-23~\unit{mV}$ yields dephasing due to electric-field fluctuations, leading to a reduced depth of the CPT dip and a general broadening of the dR spectrum (Fig.~\ref{CPT}c). We find that the electric-field fluctuations, which are probably due to rapid filling and emptying of charge traps around the QD, are reduced by illuminating the sample with a weak non-resonant ($850~\unit{nm}$) laser~\cite{Houel:2011}; switching this laser off thus results in a reduced CPT dip even quite close to $V_0$ (Fig.~\ref{CPT}e). Most strikingly, by tuning the magnetic field to a value below that of typical nuclear Overhauser fields ($B_n \sim 20~\unit{mT}$), we find that the single strong dark resonance turns into two shallow transparency dips (Fig.~\ref{CPT}d). In this regime the in-plane component of $B_n$ ensures that $\mathrm{T_{+}}$ and $\mathrm{T_{-}}$ gain some $\mathrm{T_{0}}$ character, enabling the formation of two extra quasi-dark states, which are no longer immune to slow Overhauser field fluctuations; in Fig.~\ref{CPT}d the leftmost of the resulting three dips is obscured due to the small but finite detuning of the pump laser. Applying a large enough external in-plane magnetic field would fully suppress the $T_0$ character of the middle one of the three modified T states~\cite{Tureci:2007}, yielding two dark resonances with a controllable splitting.

\begin{figure}[t]
\includegraphics[width=63.4mm]{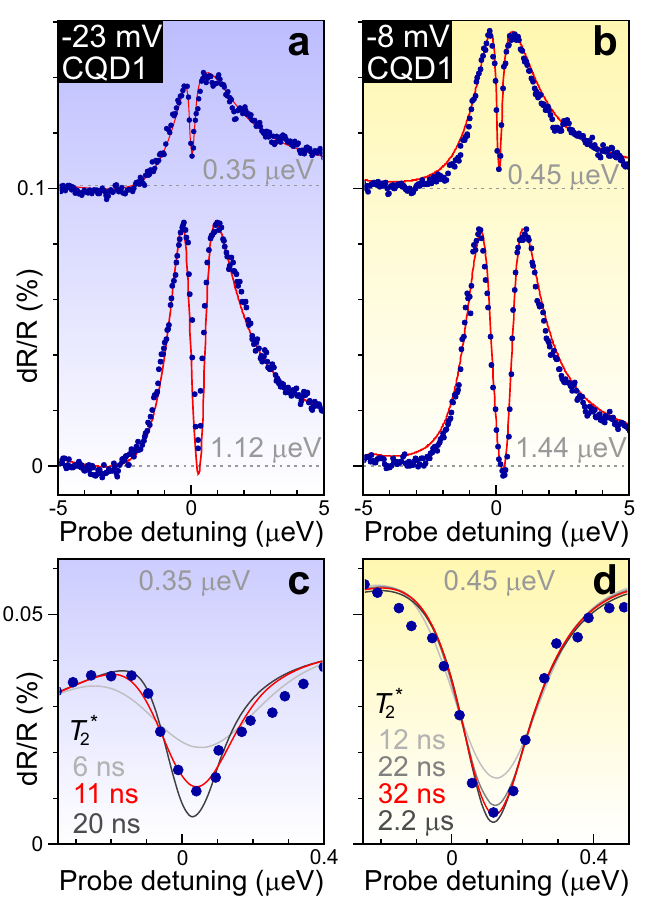}
\caption{(color online). Quantitative analysis of the $T_2^*$ decoherence time in CQD1 at $0.2 \unit{T}$. (a) dR spectra (blue dots) far away from the sweet spot ($\Delta V = -23 \unit{mV}$), for two different pump Rabi frequencies $\Omega_{\mathrm{T_0}}$ (indicated in gray). Numerical fits~$(27)$ to the data are shown in red. Traces are offset vertically for clarity. (b) dR spectra and fits closer to the sweet spot ($\Delta V = -8~ \unit{mV}$). (c) Close-up of the upper trace in (a). The spectrum is fitted with different values of $\Delta V$ while keeping $T_2^{tunnel}=250~\unit{ns}$ constant. From light gray to red to black: $\Delta V=-43, -23, -13~ \unit{mV}$, corresponding to the $T_2^*$ values indicated in the figure. (d) Close-up of the upper trace in (b), fitted with $\Delta V=-22, -12, -8, 0~\unit{mV}$.
}
\label{T2star}
\end{figure}

To quantify the coherence time we measure the CPT dip for different pump laser powers and gate voltage detunings (Fig.~\ref{T2star}). The results are then analyzed by numerically solving the optical Bloch equations ~\cite{supplemental} for the full eight-level system shown in Fig.~\ref{Scheme}b. We find that there are two mechanisms that determine the depth of the CPT dip. Far away from the sweet spot (Fig.~\ref{T2star}a) the coherence is limited by Gaussian charge fluctuations (with standard deviation $\delta V=0.6~\unit{mV}$ and long decorrelation time) that lead to fluctuations in $E_{ST}$, resulting in a spin dephasing time $T_2^{*}=11~\unit{ns}$ (Fig.~\ref{T2star}c). Moving closer towards the sweet spot (Fig.~\ref{T2star}b), the effect of the charge fluctuations becomes weaker. However, spin-flip tunneling with the back contact~\cite{Elzerman:2011} now becomes stronger, since in this coupled QD pair the sweet spot is located very close to the edge of the (1,1) regime. We capture this in our model by including an additional markovian spin dephasing term $T_2^{tunnel}=250~\mathrm{ns}$, leading to $T_2^{*}=32~\unit{ns}$ (see Fig.~\ref{T2star}d). From our quantitative
understanding of the relevant decoherence processes we can extrapolate that in the absence of (co-)tunneling, $T_2^*$ at the sweet spot should well exceed 1 microsecond, limited by second-order charge fluctuations. However, second-order hyperfine processes (not included in the simulations) would in this case limit the achievable
coherence time to $T_2^*\sim 1~\unit{\mu s}$. 

\begin{figure}[t]
\includegraphics[width=64mm]{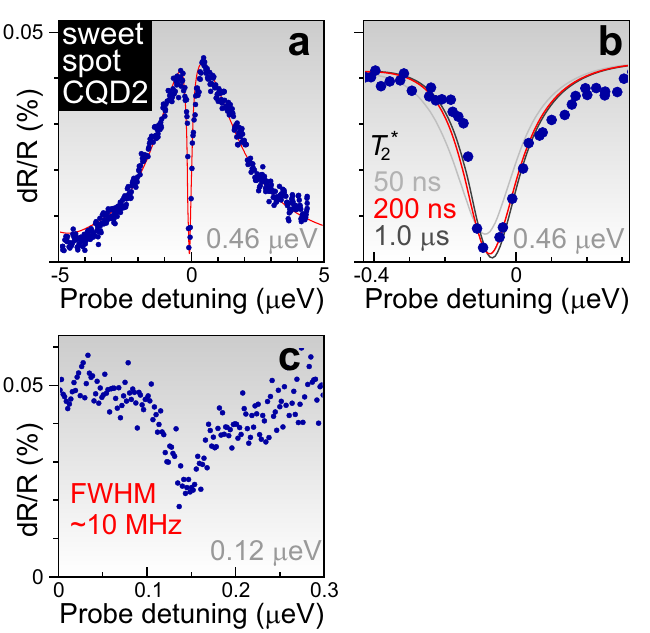}
\caption{(color online). Suppressed decoherence close to the sweet spot in CQD2. (a) dR spectra (blue dots) with $\Omega_{\mathrm{T_0}}=0.46~\unit{\mu eV}$ at $B=0.2~\unit{T}$. Due to sample drifts, tuning the gate voltage exactly to the sweet spot is challenging, so we estimate that $\Delta V = 0 \pm 2~\unit{mV}$. (b) Close-up of (a). The numerical fits~$(27)$ correspond to the $T_2^*$ values indicated in the figure, with $T_2^*=200~\unit{ns}$ fitting best. (c) CPT spectrum for very low pump and probe power ($\Omega_{\mathrm{S}}\approx \Omega_{\mathrm{T_0}}= 0.12~\unit{\mu eV}$). In this regime, the CPT dip has a full-width at half-maximum of only $\sim 10~\unit{MHz}$. 
}
\label{CQD2}
\end{figure}

To demonstrate such long $T_2^*$ times, we find another coupled QD pair where the sweet spot is  further away from the edge of the (1,1) plateau, so that tunneling-induced spin dephasing is strongly suppressed. For this second QD molecule (CQD2), our fitting procedure yields $T_2^*\geq200~\unit{ns}$, limited by the finite noise floor of our measurements (Fig.~\ref{CQD2}a and b). This lower bound on $T_2^*$ is more than two orders of magnitude longer than previously reported values for coupled electron~\cite{Kim:2010} or hole~\cite{Greilich:2011} spins away from the sweet spot; in addition, it is more than an order of magnitude longer than $T_2^{*}$ for a single electron~\cite{Hanson:2007,Mikkelsen:2007,Petta:2005,Press:2010}, and comparable to that of a single hole~\cite{Brunner:2009}. Our system thus maintains coherence on timescales that previously required spin echo
techniques; the corresponding reduction in overhead can be very beneficial for applications in quantum information processing. Conversely, the long $T_2^{*}$ should improve the effectiveness of a spin echo pulse, and could thus lead to even longer spin echo $T_2$ times. Finally, the potential of our system for high-resolution spectroscopy is highlighted in Fig.~\ref{CQD2}c; reducing the pump and probe Rabi frequencies to $\Omega_{\mathrm{S}}\approx\Omega_{\mathrm{T_0}}=0.12~\mu \mathrm{eV}$ yields a narrow CPT dip with a full width at half maximum of just $\sim 10 \unit{MHz}$. 

In addition to featuring a decoherence-avoiding qubit that can be robust against both electric and magnetic fluctuations, the two-electron CQD molecule offers additional useful features. We find that in general the electronic g-factors in each of the two coupled dots are slightly different ($g_e^{red}=0.53$ and $g_e^{blue}=0.47$), which detunes the $\Tp - \Rpp$ from the $\To-\Rp$ transition, allowing them to be separately addressed at moderate magnetic fields (see supplemental Fig.~S5 a and b). To implement single-shot spin read-out~\cite{Vamivakas:2010}, which requires recycling transitions, the S-population could be directly transferred to the $\Rpp$ state with a strong laser, and subsequently read out using light scattering on the $\Tp-\Rpp$ transition. In this sense, the rich optical excitation spectrum of QD molecules in the (1,1) regime thus combines the advantages of both Voigt~\cite{Xu:2007} and Faraday~\cite{Yilmaz:2010} geometries. 

Another very interesting possibility is highlighted in Fig.~\ref{CPT}d where it can be seen that application of an in-plane magnetic field yields two dark resonances~\cite{Tureci:2007}: it has been shown theoretically~\cite{Moeller:2007} that by adiabatically changing the
laser intensity and phase in a three-laser geometry, it is possible to realize a Hadamard Berry-phase gate, rotating the system wave-function from one dark state to a coherent superposition of the two dark states.
\\
\\

This work is supported by the Swiss National Science Foundation NCCR Quantum Photonics project and an ERC Advanced
Investigator grant (A. I.).

\newpage
\onecolumngrid
\begin{center}
\textbf{SUPPLEMENTAL MATERIAL}\\
\textbf{Decoherence-avoiding spin qubits in optically active quantum dot molecules}\\
K. M. Weiss, J. M. Elzerman, Y. L. Delley, J. Miguel-Sanchez, and A. Imamo\u{g}lu\\
\textit{Institute of Quantum Electronics, ETH Zurich, CH-8093 Zurich, Switzerland.}\\
\end{center}


\section{Device and experimental setup}
The vertically stacked InGaAs QDs are embedded in a Schottky diode, grown by molecular beam epitaxy (MBE) on a (100) GaAs substrate. The QDs in the second layer tend to nucleate directly on top of the first layer, due to the strain field~\cite{Xie:1995} produced by the latter. The partially-covered-island technique is employed  to reduce the thickness of the QDs and thereby blue shift their emission wavelength into the near-infrared ($\sim 940-970~\unit{nm}$). The wavelength of the top (red) QD is designed to be $\sim 30~\unit{nm}$ larger than the bottom (blue) QD wavelength. This, together with choosing the interdot tunnel barrier to be $9~\unit{nm}$, allows for charging each QD with one electron by applying an appropriate bias voltage $V$.

This bias voltage is applied between the Si-doped $n^+$-GaAs back contact ($50~\unit{nm}$ below the bottom QD layer) and a semitransparent top gate ($2~\unit{nm}$ of Ti plus $6~\unit{nm}$ of Au), evaporated after growth. By tuning $V$, we can controllably charge the QDs and shift their electronic energy levels into resonance, enabling tunneling. To reduce current through the device a $40~\unit{nm}$ AlGaAs blocking layer is grown $218~\unit{nm}$ above the top QD-layer.

The device is mounted on a three-axis piezoelectric nano-positioning stack in a liquid-helium bath cryostat, operating at $4.2~\unit{K}$. A lens with a numerical aperture of $0.68$ focuses the incoming light onto the sample to a near-diffraction limited spot, which allows for addressing a single CQD pair. Resonant differential transmission (dT) and reflection (dR) measurements in combination with Stark-shift modulation spectroscopy~\cite{Hoegele:2004} are employed. For dR, the scattered light is collected with the focusing lens and detected with a Si photodiode at room temperature. Therefore, it is possible to use linear polarizers to block a strong coupling laser, cross-polarized with respect to the polarizer before the detector.

For most measurements, a weak non-resonant laser operating at $850~\unit{nm}$ is focused on the device. We find that this reduces the charge fluctuations in the sample, resulting in narrower and stronger dT and dR resonances and a deeper CPT dip.

\section{Theoretical model and numerical fitting procedure}
We model our two-electron coupled QD by taking into account all four S and T ground states plus the four optically excited states R for which the red QD contains a hole (see Fig.~1b in the main text). We then solve the corresponding eight-level master equation in the steady-state. Decoherence processes with a fast decorrelation time (such as the spin-flip tunneling with the back contact mentioned in the main text) are included in the Lindblad form. Decoherence processes with a slow decorrelation time (such as charge fluctuations that correspond to a fluctuating voltage $V$) are included as follows. First, we assume a particular value for $V$ and calculate the corresponding value of $E_{ST}$ (from a fit of $E_{ST}$ versus $V$, as shown in the inset to Fig.~1d in the main text). Then we numerically solve the master equation with this particular value of $E_{ST}$ to find the corresponding steady-state solution of dR versus probe laser detuning. This trace is weighed with a factor corresponding to a Gaussian distribution of a known mean (i.e. the gate voltage we apply) and a known standard deviation $\delta V$ (which we have determined from an independent measurement). This procedure is then repeated for another particular value of the gate voltage. Finally, all traces (100 in total) are averaged with their proper Gaussian weighing factor. 

The above procedure gives us both the absorptive as well as the dispersive part of the QD response. To account for the observed asymmetry in the measured dR traces, we add $\sim 15\%$ of the dispersive part to the absorptive part of the QD response. This mimics a well-known optical interference effect~\cite{Karrai:2003} that gives rise to an asymmetry that is more pronounced in reflection than in transmission measurements (see supplementary Fig.~S1). By fitting the simulated trace to the measured data, we obtain values for the fitting parameters $\Delta V=V-V_0$ (i.e. the distance to the sweet spot) and $T_2^{markov}$ (i.e. the total markovian decoherence rate between $\mathrm{S}$ and $\To$). (We treat $\Delta V$ as a fitting parameter, since it is very hard to determine it experimentally due to long-term drifts and instabilities in the device.)

Finally, from the values of the fit parameters $T_2^{markov}$ and $\Delta V$ (in combination with the independently determined value of $\delta V$), we can determine the corresponding value of $T_2^{*}$. As before, we first assume a particular value of $V$ (corresponding to a particular $E_{ST}$) and calculate the corresponding damped harmonic oscillation that would be seen in a time-resolved measurement of the Ramsey fringes. This time-trace is then weighed with a factor corresponding to the Gaussian distribution of $V$, and the procedure is repeated for another particular value of $V$. Finally all traces are averaged with their proper Gaussian weighing factor. $T_2^*$ then corresponds to the time it takes until this ``inhomogeneously broadened" coherence has decayed to $1/e$ of its initial value.

\begin{figure}[t]
\centering
\includegraphics{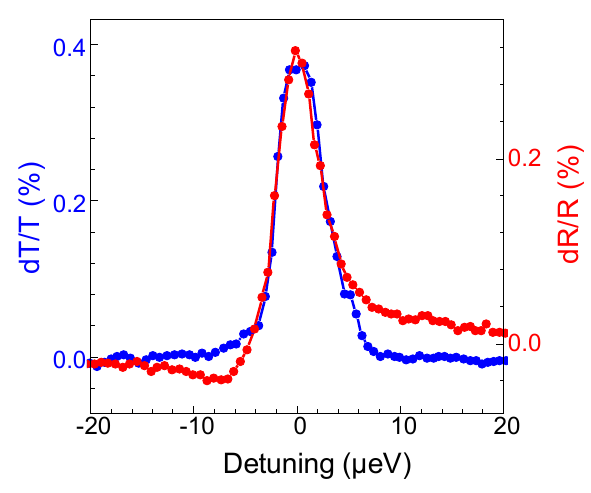}\\
\begin{flushleft}
\textbf{FIG.~S1.} Asymmetric line shape due to optical interference. Differential transmission (dT; blue trace) and differential reflection (dR; red trace) measurements of the neutral exciton in the red QD at $B=0~\unit{T}$ with laser power below saturation ($\Omega=0.34~\unit{\mu eV}$). The two traces were taken for the same optical alignment, but show very different asymmetry, indicating an optical interference effect~\cite{Karrai:2003}.
\end{flushleft}
\end{figure}

Below, we detail the procedure by which we determine the parameters used in the numerical simulations. First, we deduce the standard deviation $\delta V$ of the typical Gaussian fluctuations for both coupled QD pairs discussed in the main text. To this end, we use the very steep line observed in the (1,2) regime (see supplementary Fig.~S2a and the top right-hand corner of Fig.~1d in the main text). This line is associated with an indirect transition, i.e. an electron in QD-B recombining with a hole in QD-R. Due to its very large dipole size ($\sim 8~\unit{nm}$), this transition is very susceptible to charge noise, giving it a Gaussian lineshape. We can reproduce this lineshape by setting $\delta V=0.6~\unit{mV}$ for CQD1 (in the presence of a weak non-resonant laser at $850~\unit{nm}$), and $\delta V=0.8~\unit{mV}$ for CQD2.

\begin{figure}[b]
\centering 
\includegraphics{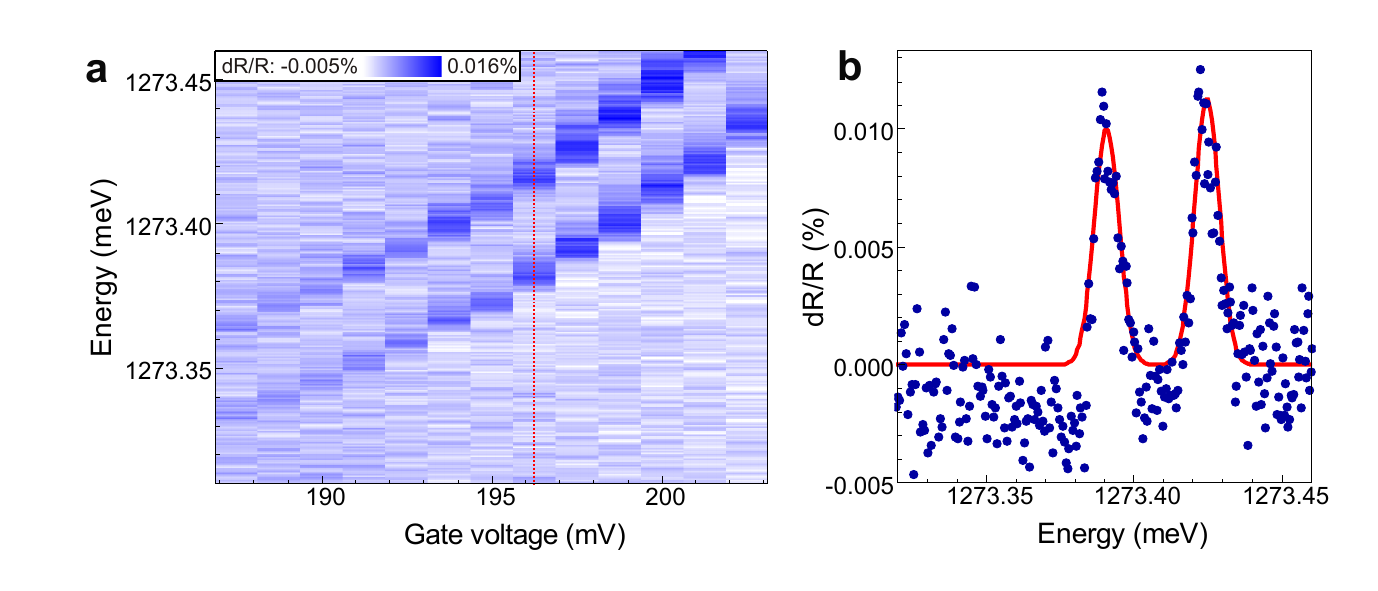}\\
\begin{flushleft}
\textbf{FIG.~S2.} Gaussian fluctuations. (a) dR measurement of the indirect transition in the (1,2)-regime versus gate voltage at $B=0.2~\unit{T}$. (b) Line cut of (a) at $196.2~\unit{mV}$, as indicated by the red line in (a). The red line in (b) is a Gaussian fit with standard deviation of the Gaussian distribution $\delta V=0.6~\unit{mV}$.
\end{flushleft}
\end{figure}

\begin{figure}[t]
\centering 
\includegraphics{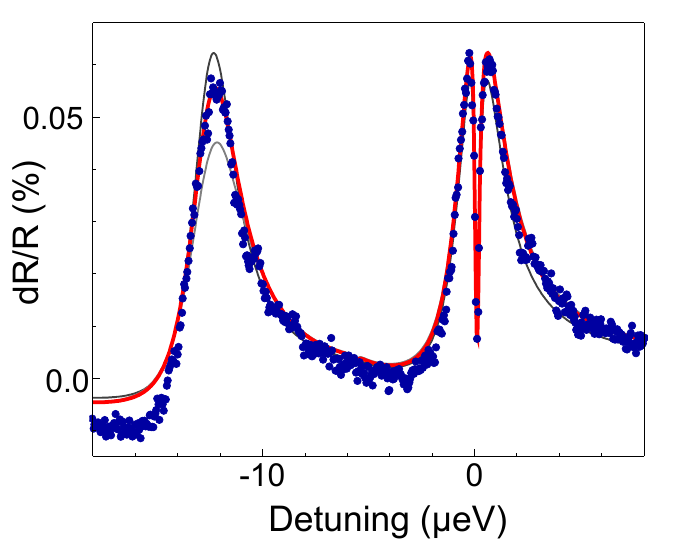}\\
\begin{flushleft}
\textbf{FIG.~S3.} Co-tunneling rates and excited state mixing. Blue dots: dR-spectrum at $B=0.2~\unit{T}$ with $\Omega_{T_0}=0.45~\unit{\mu eV}$. Red line: fits employing the eight-level simulation with $\gamma_g=2.5\cdot10^{-3}~\unit{\mu eV}$ and $\gamma_{es}=2.5\cdot10^{-2}~\unit{\mu eV}$. Dark gray line: $\gamma_g=5\cdot10^{-3}~\unit{\mu eV}$ and $\gamma_{es}=2.5\cdot10^{-2}~\unit{\mu eV}$; light gray line: $\gamma_g=2.5\cdot10^{-3}~\unit{\mu eV}$ and $\gamma_{es}=5\cdot10^{-2}~\unit{\mu eV}$.
\end{flushleft}
\end{figure}

\begin{figure}[b]
\centering 
\includegraphics{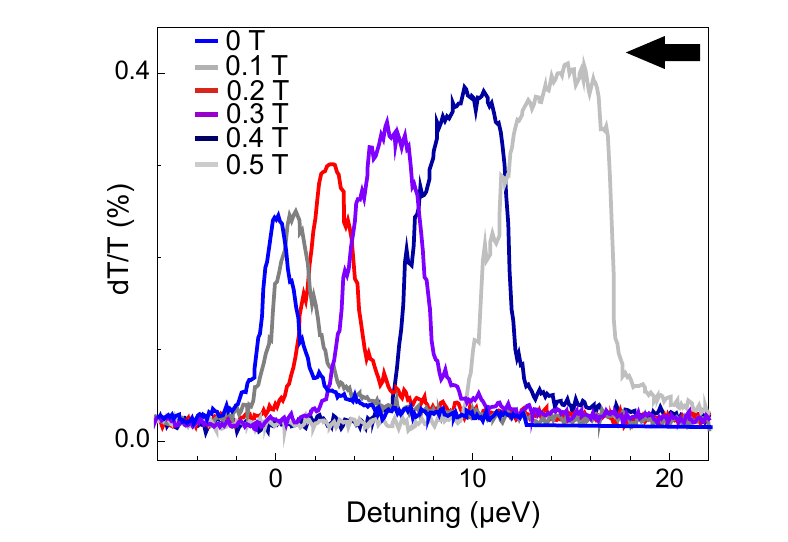}\\
\begin{flushleft}
\textbf{FIG.~S4.} Nuclear spin effects. dT measurement of blue Zeeman transition of the neutral exciton in QD-R for various B-fields: $B=0~\unit{T}$ to $B=0.5~\unit{T}$. The black arrow indicates the sweep direction of the laser.
\end{flushleft}
\end{figure}

The Lorentzian ground state dephasing processes result from the sum of all rates into and out of S, stemming from gate voltage dependent (co-)tunneling to the back contact~\cite{Smith:2005}. Due to the large value of the thermal energy of the back contact compared to $E_{ST}$, we can assume the total rate into and out of S to be equal, and we denote it $\gamma_g$. Tunneling between the excited states (i.e. between $\Rmm$ and $\Rm$ as well as between $\Rpp$ and $\Rp$) is included with a rate $\gamma_{es}$. The rates $\gamma_g$ and $\gamma_{es}$ can be determined by the contrast of the S to $\Rp$ and S to $\Rm$ transitions (shown in Fig.~S3). For strong excited-state mixing (light gray line), the contrast of the S to $\Rm$ transition is underestimated, whereas it is overestimated for strong ground state mixing (dark gray line). 

Another factor we take into account is the ``branching ratio" between optically allowed transitions (e.g. $\Rpp$ to $\Tp$) and optically forbidden ones (e.g. $\Rpp$ to $\To$ and S). For single QDs the branching ratio has been determined experimentally to be $\sim 250$, and we assume the branching ratio to be the same for our coupled QDs.

To reproduce the overall width of the optical spectra (Figs. 2 and 3 in the main text) an additional Lorentzian excited state dephasing rate $\gamma$ is introduced. We cannot completely rule out effects of the nuclear spins on the lineshape; for instance, it can be seen in Fig.~S4 that the lineshape of the neutral exciton at $B=0.2~\unit{T}$ clearly deviates from the one at $B=0~\unit{T}$. 

The values of the fitting parameters employed in Figs.~2, 3 and 4 of the main text are given in the table below. In order to include second order hyperfine processes at the sweets spot, leading to additional Gaussian dephasing, $s$ is increased in the simulation in Fig.~4 from $3.8\cdot10^{-3}~\unit{\mu eV/{mV^2}}$ to $1.9\cdot10^{-2}~\unit{\mu eV/{mV^2}}$.
\begin{table}[h!]
	\centering
		\begin{tabular}{c|c|c|c|c|c|c}
			Fig. & $\Delta V$ (mV) & $\delta V$ (mV) & $s$ ($\mathrm{\mu eV/{mV^2}}$) & $\Gamma_{\unit{S}}=\Gamma_{\unit{T_0}}$ ($\mu\mathrm{eV}$) & $\gamma$ ($\mu\mathrm{eV}$) & $1/\gamma_g$ (ns)\\ \hline
			2b, 3b  & -8  & 0.6 & $2.99\cdot10^{-3}$  & 0.4 & 0.4 & 250\\ 
			2c, 3a  & -23 & 0.6 & $2.99\cdot10^{-3}$  & 0.4 & 0.6 & 1300\\
			2d  & -13 & 0.6 & $2.99\cdot10^{-3}$  & 0.4 & 0.4 & 400\\
			2e  & -8  & 1.5 & $2.99\cdot10^{-3}$  & 0.4 & 0.4 & 250\\\hline
			4a, b  & 0 & 0.8 & $1.9\cdot10^{-2}$  & 0.4 & 0.47 & 2600\\
		\end{tabular}
		\caption{Values of the parameters used to fit the data in the main text.}
\end{table}

To determine the source of the electrical fluctuations, we compare resonance fluorescence (RF) measurements to dR measurements. The RF measurements are taken while strongly filtering the applied gate voltage, using a $1~\unit{Hz}$ low-pass filter at room temperature. The dR measurements do not allow for such low-frequency filtering due to the applied voltage modulation of $100~\unit{mV}$ at $1.3~\unit{kHz}$. Nevertheless, we find no significant difference in the depth of the CPT dip or the overall width of the spectra for the two different types of measurements. Hence, we believe the source of the electrical fluctuations is in the device itself, most likely resulting from charged defects in the QD environment.   

\section{Quasi-recycling transition}

\begin{figure}[t]
\centering 
\includegraphics{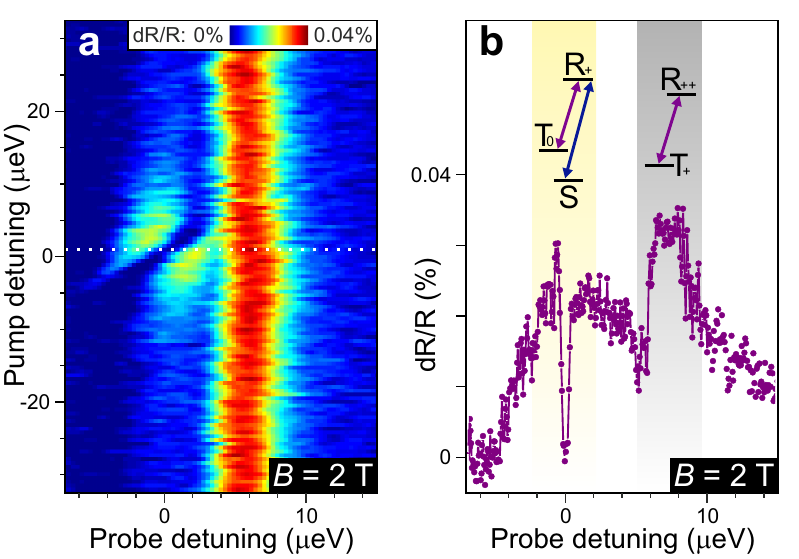}\\
\begin{flushleft}
\textbf{FIG.~S5.} Quasi-recycling transition. (a) Probing the $\To$ to $\Rp$ and $\Tp$ to $\Rpp$ transitions with $\Omega_{\mathrm{T}}=0.34~\unit{\mu eV}$ while pumping the $\mathrm{S}$ to $\Rp$ transition with $\Omega_{\mathrm{S}}=3.4~\unit{\mu eV}$ at $B=2~\unit{T}$ and $\Delta V=-15~\unit{mV}$. In contrast to the $\Tp$ to $\Rpp$ transition, the $\To$ to $\Rp$ transition is only visible at the pump laser resonance. (b) CPT spectrum of the $\To$ to $\Rp$ transition at $B=2~\unit{T}$ and $\Delta V=-11~\unit{mV}$ with $\Omega_{\mathrm{S}}=1.06~\unit{\mu eV}$. At blue detuning the quasi-recycling $\Tp$ to $\Rpp$ transition becomes visible.
\end{flushleft}
\end{figure}

\begin{figure}[b]
\centering 
\includegraphics{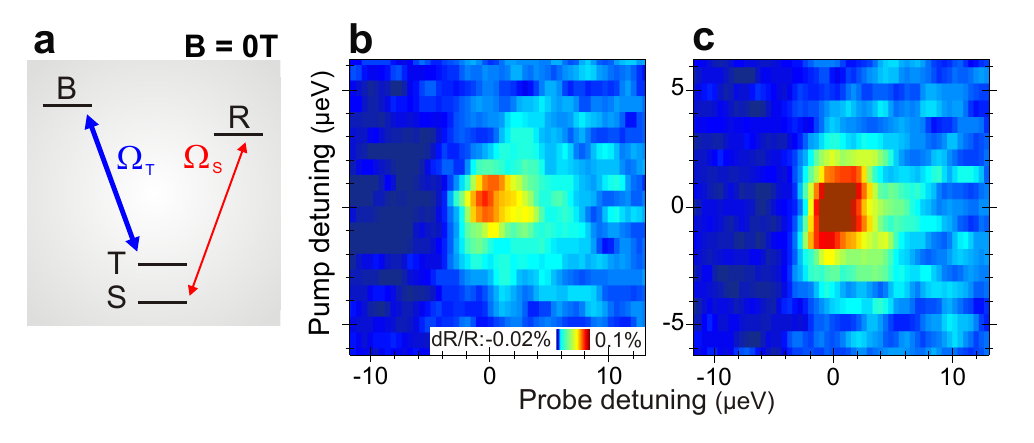}\\
\begin{flushleft}
\textbf{FIG.~S6.} Double $\Lambda$-scheme in the (1,1) regime. (a) Schematic of the double $\Lambda$-scheme, formed by S and T ground states together with the optically excited states in the red and blue QD. A stronger repump laser is applied to the T to B transition of the blue QD and a probe laser to the S to R transitions of the red QD. The corresponding experiment is shown in (b), performed in dR with $\Omega_{coupling}=0.68~\unit{\mu eV}$ and $\Omega_{probe}=2.0~\unit{\mu eV}$ at $B=0~\unit{T}$ and $\Delta V=-93~\unit{mV}$. (c) The probe laser is now swept across the T to R transitions of the red QD, while stepping the coupling laser over the S to B transition in the blue QD. The contrast is therefore three times higher than in (b).
\end{flushleft}
\end{figure}

As mentioned in the main text, the rich optical excitation spectrum of QD molecules in the (1,1) regime combines the advantages of both Voigt and Faraday geometries. This will now be explained in more detail. Spin manipulation protocols require $\Lambda$-schemes with comparable oscillator strength on both transitions, as one can find in single negatively charged QDs in Voigt geometry. In our system, such a $\Lambda$-system is formed by the $\To$ and $\mathrm{S}$ ground states coupled to $\Rp$. For single-shot spin read-out, recycling transitions that one obtains in Faraday geometry are needed. The two-electron QD molecule exhibits two such recycling ($\mathrm{T_{\pm}}$ to $\mathrm{R_{\pm \pm}}$) transitions at moderate magnetic fields, whenever the electron g-factors in the two QDs differ. We find that this is precisely the case for our QD-molecule where the g-factors differ by 0.06: Fig.~S5 shows that for $B = 2~\unit{T}$, the transition $\Tp$-$\Rpp$ is clearly detuned from the $\To$-$\Rp$ transition exhibiting the transparency dip. 

\section{Double-lambda scheme}
A parametric amplifier based on a double lambda scheme for instance, can be implemented by coupling the excited states in the red and blue QD to the $\mathrm{S}$-$\To$ ground states~\cite{Lukin:1998,Jain:1996}. To verify that the excited states of both QDs share common ground states, we perform repump experiments (Figs.~S6b and c). To this end we probe the S (T) transition in the red QD while pumping the T (S) transition in the blue QD. When both lasers are on resonance, the spin state is randomized, leading to a sizable dR signal. Another double $\Lambda$-scheme is formed at finite magnetic fields. Here, the S and T ground states couple to $\Rp$ and $\Rm$ in the red QD ($\mathrm{B_+}$ and $\mathrm{B_-}$ in the blue QD). The emission energies for both types of double $\Lambda$-schemes are highly tunable by adjusting either growth parameters (QD size, $\Delta E_{\mathrm{ST}}$ via interdot tunnel barrier) or the applied magnetic field and bias voltage.\\


\begin{thebibliography}{}
\bibitem{Hanson:2007}
R. Hanson, L. P. Kouwenhoven, J. R. Petta, S. Tarucha, and L. M. K. Vandersypen, \textit{Rev. Mod. Phys.} \textbf{79}, 1217 (2007).

\bibitem{Mikkelsen:2007}
Mikkelsen, J. Berezovsky, N. G. Stoltz, L. A. Coldren, and D. D.  Awschalom, \textit{Nature Phys.} \textbf{3}, 770-773 (2007).

\bibitem{Press:2008}
D. Press, T. D. Ladd, B. Zhang, and Y. Yamamoto, \textit{Nature} \textbf{456}, 218-221 (2008).

\bibitem{Xu:2008}
X. Xu \textit{et al.}, \textit{Nature Phys.} \textbf{4}, 692-695 (2008).

\bibitem{Brunner:2009}
D. Brunner \textit{et al.}, \textit{Science} \textbf{325} 70-72 (2009).

\bibitem{DeGreve:2011}
K. De Greve \textit{et al.}, \textit{Nature Phys.} \textbf{7}, 872-878 (2011).

\bibitem{Petta:2005}
J. R. Petta \textit{et al.}, \textit{Science} \textbf{309}, 2180-2184 (2005).

\bibitem{Bluhm:2010}
H. Bluhm \textit{et al.}, \textit{Nature Phys.} \textbf{7}, 109-113 (2010).

\bibitem{Press:2010}
D. Press \textit{et al.}, \textit{Nature Phot.} \textbf{4}, 367-370 (2010).

\bibitem{Latta:2009}
C. Latta \textit{et al.}, \textit{Nature Phys.} \textbf{5}, 758-763 (2009).

\bibitem{Xu:2009}
X. Xu \textit{et al.}, \textit{Nature Phys.} \textbf{5}, 1105-1109 (2009).

\bibitem{BluhmPRL:2010}
H. Bluhm, S. Foletti, D. Mahalu, V. Umansky, and A. Yacoby, \textit{Phys. Rev. Lett.} \textbf{105}, 216803 (2010).

\bibitem{Lidar:1998}
D. A. Lidar, I. L. Chuang, and K. B. Whaley, \textit{Phys. Rev. Lett.} \textbf{81}, 2594-2597 (1998).

\bibitem{Kim:2010}
D. Kim, S. C. Carter, A. Greilich, A. S. Bracker, and D. Gammon, \textit{Nature Phys.} \textbf{4}, 223-228 (2010).

\bibitem{Elzerman:2011}
J. M. Elzerman, K. M. Weiss, J. Miguel-Sanchez, and A. Imamoglu, \textit{Phys. Rev. Lett.} \textbf{107}, 017401 (2011).

\bibitem{Greilich:2011}
A. Greilich, S. G. Carter, D. Kim, A. S. Bracker, and D. Gammon, \textit{Nature Phot.} \textbf{5}, 702-708 (2011).

\bibitem{Stinaff:2006}
E. A. Stinaff \textit{et al.}, \textit{Science} \textbf{311}, 636-639 (2006).

\bibitem{Tureci:2007}
H. E. T\"ureci, J. M. Taylor, and A. Imamoglu, \textit{Phys. Rev. B} \textbf{75}, 235313 (2007).

\bibitem{Vion:2002}
D. Vion \textit{et al.}, \textit{Science} \textbf{296}, 886 (2002).

\bibitem{Koch:2007}
J. Koch \textit{et al.}, \textit{Phys. Rev. A} \textbf{76}, 042319 (2007).

\bibitem{Vanier:2005}
J. Vanier, \textit {Appl. Phys. B}, \textbf{81}, 421-442  (2005).

\bibitem{Krenner:2005}
H. J. Krenner \textit{et al.}, \textit{Phys. Rev. Lett.} \textbf{94}, 057402 (2005).

\bibitem{supplemental}
See Supplemental Material for details of the materials, methods and theoretical analysis.

\bibitem{Alen:2003}
B. Al\'en, F. Bickel, K. Karrai, R. J. Warburton, and P. M. Petroff, \textit{Appl. Phys. Lett.} \textbf{66}, 2593-2596 (2003).

\bibitem{Fleischhauer:2005}
M. Fleischhauer, A. Imamoglu, and J. P. Marangos, \textit{Rev. Mod. Phys.} \textbf{77}, 633 (2005).

\bibitem{Boller:1991}
K.-J. Boller, A. Imamoglu, and S. E. Harris, \textit{Phys. Rev. Lett.} \textbf{66}, 2593 (1991).

\bibitem{Houel:2011}
J. Houel \textit{et al.}, \textit{arXiv:1110.2714} (2011).

\bibitem{Vamivakas:2010}
A. N. Vamivakas \textit{et al.}, \textit{Nature} \textbf{467}, 297-300 (2010).

\bibitem{Xu:2007}
X. Xu \textit{et al.}, \textit{Phys. Rev. Lett.} \textbf{99}, 097401 (2007).

\bibitem{Yilmaz:2010}
S. T. Yilmaz, P. Fallahi, and A. Imamoglu, \textit{Phys. Rev. Lett.} \textbf{105}, 033601 (2010).

\bibitem{Moeller:2007}
D. M\o{}ller, L. B. Madsen, and K. M\o{}lmer, \textit{Phys. Rev. A} \textbf{75}, 062302 (2007).

\end{thebibliography}

\begin{thebibliography}{}

\bibitem{Xie:1995}
Q. Xie, A. Madhukar, P. Chen, and N. P. Kobayashi, \textit{Phys. Rev. Lett.} \textbf{75}, 2542 (1995).

\bibitem{Hoegele:2004}
A. H\"ogele \textit{et al.}, \textit{Phys. Rev. Lett.} \textbf{93}, 217401 (2004).

\bibitem{Karrai:2003}
K. Karrai and R. J. Warburton, \textit{Superlattices and Microstructures} \textbf{33}, 311–337 (2003).

\bibitem{Smith:2005}
J. M. Smith \textit{et al.}, \textit{Phys. Rev. Lett.} \textbf{94}, 197402 (2005).

\bibitem{Lukin:1998}
M. D. Lukin, P. R. Hemmer, M. L\"offler, and M. O. Scully, \textit{Phys. Rev. Lett.} \textbf{81}, 2675 (1998).

\bibitem{Jain:1996}
M. Jain, H. Xia, G. Y. Yin, A. J. Merriam, and S. E. Harris, \textit{Phys. Rev. Lett.} \textbf{77}, 4326 (1996).


\end{thebibliography}
\end{document}